\documentclass[aps,prl,10pt,final,letterpaper,twocolumn]{revtex4}

\usepackage{graphicx}
\usepackage{amssymb,amsfonts,amsmath}
\usepackage{float}

\makeatletter
\def\@dotsep{4.5}
\makeatother

\newcommand{\id}{\mathrm{d}} 
\newcommand{\h}{\mathcal{H}}
\newcommand{\E}{\mathcal{E}}

\newlength{\colwidth}
\begin{document}

\title{Inverted Cheerios effect: Liquid drops attract or repel by elasto-capillarity}
\date{\today}

\author{S. Karpitschka$^1$}
\author{A. Pandey$^1$}
\author{L.A. Lubbers$^2$}
\author{J.H. Weijs$^3$}
\author{L. Botto$^4$}
\author{S. Das$^5$}
\author{B. Andreotti$^6$}
\author{J.H. Snoeijer$^{1,7}$}
\affiliation{$^1$Physics of Fluids Group, Faculty of Science and Technology, University of Twente, P.O. Box 217, 7500 AE Enschede, The Netherlands\\
$^2$Huygens-Kamerlingh Onnes Lab, Universiteit Leiden, P.O. Box 9504, 2300 RA Leiden, The Netherlands\\ 
$^3$Laboratoire de Physique de l{'}{\'E}cole Normale Sup{\'e}rieure de Lyon, Universit{\'e} de Lyon, 46, all{\'e}e d{'}Italie, 69007 Lyon, France\\
$^4$School of Engineering and Materials Science, Queen Mary University of London, London E1 4NS, UK\\
$^5$Department of Mechanical Engineering, University of Maryland, College Park, MD 20742, USA\\
$^6$Laboratoire de Physique et M{\'e}canique des Milieux Het{\'e}rog{\`e}nes, UMR 7636 ESPCI- CNRS, Universit{\'e} Paris- Diderot, 10 rue Vauquelin, 75005 Paris, France\\
$^7$Department of Applied Physics, Eindhoven University of Technology, P.O. Box 513, 5600MB
Eindhoven, The Netherlands.
}

\begin{abstract} 	
Solid particles floating at a liquid interface exhibit a long-ranged attraction mediated by surface tension. In the absence of bulk elasticity, this is the dominant lateral interaction of mechanical origin. Here we show that an analogous long-range interaction occurs between adjacent droplets on solid substrates, which crucially relies on a combination of capillarity and bulk elasticity. We experimentally observe the interaction between droplets on soft gels and provide a theoretical framework that quantitatively predicts the migration velocity of the droplets. Remarkably, we find that while on thick substrates the interaction is purely attractive and leads to drop-drop coalescence, for relatively thin substrates a short-range repulsion occurs which prevents the two drops from coming into direct contact. This versatile, new interaction is the liquid-on-solid analogue of the ``Cheerios effect". The effect will strongly influence the condensation and coarsening of drop soft polymer films, and has potential implications for colloidal assembly and in mechanobiology.
\end{abstract}

\maketitle


The long-ranged interaction between particles trapped at a fluid interface is exploited for the fabrication of microstructured materials via self-assembly and self-patterning \cite{tessier2001,Furst11,Cavallaro13,bowden1997,Ershov13} and occurs widely in the natural environment when living organisms or fine particles float on the surface of water \cite{loudet2011, peruzzo2013}. In a certain class of capillary interactions the particles deform the interface because of their shape or chemical heterogeneity \cite{danov2005,botto2012,kumar2013}. In this case the change in interfacial area upon particle-particle approach causes an attractive capillary interaction between the particles. In the so-called Cheerios effect, the interaction between floating objects is mainly due to the change in gravitational potential energy associated to the weight of the particles, which deform the interface while being supported by surface tension  \cite{Vella2005}, and the same principle applies when the interface is elastic \cite{Maha2015}. The name ``Cheerios effect''  is reminiscent of breakfast cereals floating on milk and sticking to each other or to the walls of the breakfast bowl . 

\begin{figure*}[t]
	\centerline{\includegraphics[width=\textwidth]{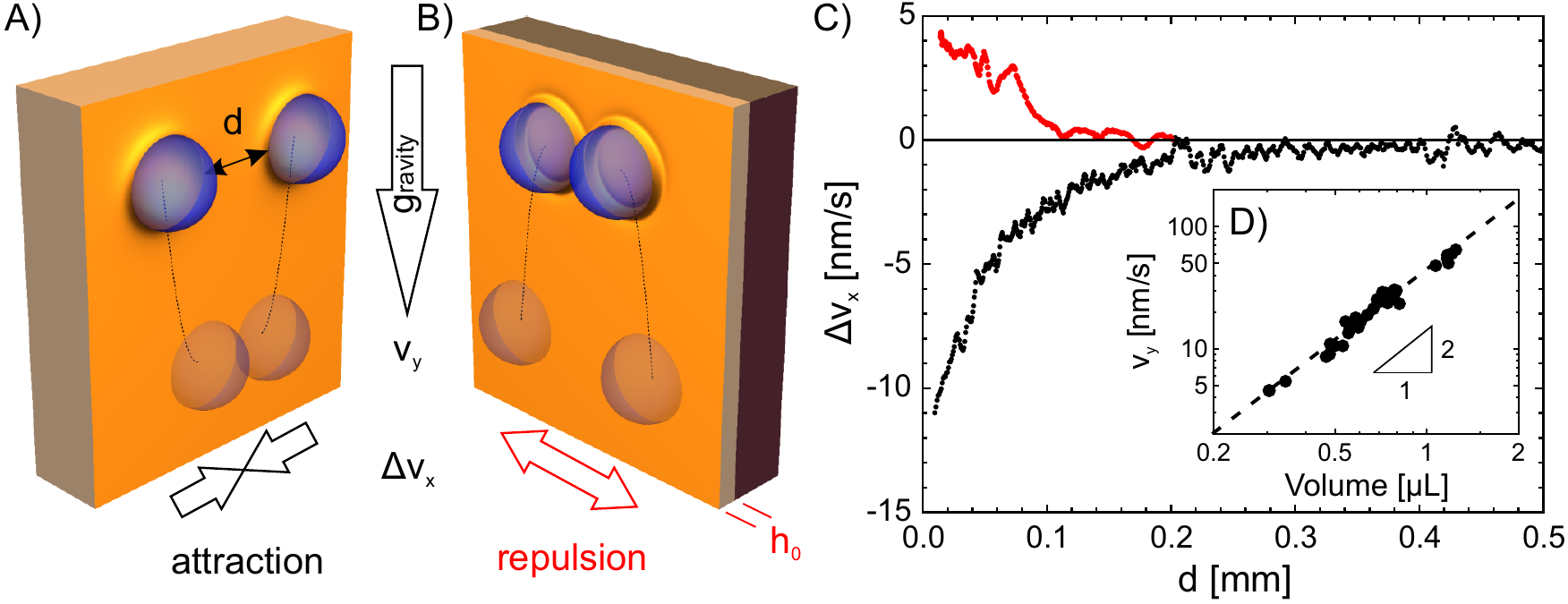}}
	\caption{The Inverse Cheerios effect for droplets on soft solids. Two liquid drops sliding down a soft gel exhibit a mutual interaction, mediated by the elastic deformation of the substrate. A) Drops sliding down a \emph{thick} elastic layer attract each other, providing a new mechanism for coalescence. B) Drops sliding down a \emph{thin} elastic layer (thickness $h_0$)  repel each other. C) Measurement of the horizontal relative velocities $\Delta v_x$ of droplet pairs, as a function of separation distance $d$. These measurement quantify the interaction strength. D) Sliding velocity of isolated droplets on the thick layer as a function of their volume. This data is used to calibrate the relation between force (gravity) and sliding velocity.}
	\label{fig1}
\end{figure*}

Here we consider a situation opposite to that of the Cheerios effect by considering liquid drops deposited on a solid. The solid is  sufficiently soft to be deformed by the surface tension of the drops, resulting in a lateral interaction.  Recent studies have provided a detailed view of statics of single-drop wetting on deformable surfaces ~\cite{JXWD11,Limat12,MDSA12b,Style12,LUUKJFM14}. The length scale over which the substrate is deformed is set by the ratio of the droplet surface tension $\gamma$ and the shear modulus $G$. The deformation can be seen as an elasto-capillary meniscus, or ``wetting ridge", around the drop (Fig.~\ref{fig1}~A,B). Interestingly, the contact angles at the edge of the drop are governed by Neumann's law, just as for oil drops floating on water.  In contrast to the statics of soft wetting, its dynamics has only been explored recently. New effects such as stick-slip induced by substrate viscoelaticity \cite{limat14,karpitschka2015} and droplet migration due to stiffness gradients \cite{durotaxisPNAS13} have been revealed. The possibility that elasto-capillarity induces an interaction between neighboring drops is of major importance for applications such as drop condensation on polymer films \cite{Sokuler09} and self-cleaning surfaces \cite{Schellenberger15,Rykaczewski14,Kim12,Subramanyam12}. The interaction between drops on soft surfaces might also provide insights into the mechanics of cell locomotion~\cite{Science2005,Lo:BPJ2000,Zhang:NatComm2014} and cell-cell interaction ~\cite{Guo:BPJ2006}.

Here we show experimentally that long-ranged elastic deformations lead to an interaction between neighboring liquid drops on a layer of cross-linked polydimethylsiloxane (PDMS). The layer is sufficiently soft for significant  surface tension induced deformations to occur (Fig.~\ref{fig1}). The interaction we observe can be thought of as the inversed Cheerios effect, since the roles of the solid and liquid phases are exchanged. Remarkably, the interaction can be either attractive or repulsive, depending on the geometry of the gel. We propose a theoretical derivation of the interaction force from a free energy calculation that self-consistently accounts for the deformability of both the liquid drop and the elastic solid.

\section{Experiment: attraction versus repulsion}
Here, the inverted Cheerios effect is observed with sub-millimeter drops of ethylene glycol on a  PDMS gel. The gel is a reticulated polymer formed by polymerizing small multifunctional prepolymers --~contrary to hydrogels, there is no liquid phase trapped inside. The low shear modulus of the PDMS gel gives an elasto-capillary length $\ell =\gamma/G =0.17\ \mathrm{mm}$ sufficiently large to be measurable in the optical domain.

The interaction between two neighboring liquid drops is quantified by tracking their positions while they are sliding under the effect of gravity along a soft layer held vertically. The interaction can be either attractive (Fig.~\ref{fig1}~A) or repulsive (Fig.~\ref{fig1}~B): drops on relatively thick gel layers attract each other, while drops on relatively thin layers experience a repulsion.

The drop-drop interaction induces a lateral motion that can be quantified by the horizontal component of the relative droplet velocity, $\Delta v_x$, with the convention that $\Delta v_x>0$ implies repulsion. In Fig.~\ref{fig1}~C, we report $\Delta v_x$ as a function of the separation $d$, defined as the shortest distance between the surfaces of the drops. The drops ($R\simeq 0.5-0.8\;\mathrm{mm}$) exhibit attraction when sliding down a thick layer ($h_0= 8\;\mathrm{mm}$, black curve), while they are repelled on a thin layer ($h_0= 0.04\;\mathrm{mm}$, red curve). $\Delta v_x$ is larger at close proximity, signaling an increase in the interaction force. Spontaneous merging occurs where drops encounter direct contact. Importantly, these interactions provide a new mechanism for droplet coarsening (or ordering) by coalescence (or its suppression) that has no counterpart on rigid surfaces.

The interaction force $F$ can be inferred from the relative velocities between the drops, based on the effective ``drag" due to sliding on a gel. We first calibrate this drag by considering drops that are sufficiently separated, so that they do not experience any mutual interaction. The motion is purely downward and driven only by gravitational force $F_g = M g$, and inertia is negligible. Figure~\ref{fig1}~D shows that the droplet velocity $v_y$ is nonlinear, and approximately scales as $F_g^2$. This force-velocity calibration curve is in good agreement with viscoelastic dissipation in the gel, based on which one expects the scaling law~\cite{karpitschka2015}:
\begin{equation}\label{eq:velocity}
v \sim \frac{\ell}{\tau} \left( \frac{F}{2\pi R\gamma}\right)^{1/n}.
\end{equation}
Here $n$ is the rheological exponent that emerges from the scale invariance of the gel network \cite{ChambonWinter,LAL96,deGennes1996}, while $\tau$ is a characteristic timescale. The parameter values $n \simeq 0.61$ and $\tau\simeq 0.68\;\mathrm{s}$ are calibrated in a rheometer (see Methods). Equation \eqref{eq:velocity} is valid for $v$ below the characteristic rheological speed, $\ell/\tau$ -- this is justified here since $\ell/\tau \sim 0.25\ \mathrm{mm/s}$ for the silicone gel, while the reported speeds reach up to $\sim 100\ \mathrm{nm/s}$. The presence of a large viscoelastic dissipation in the gel exceeds the dissipation within the drop by orders of magnitude, and explains these extremely slow drop velocities observed experimentally~\cite{CGS96,karpitschka2015}. The force-distance relation for the Inverse Cheerios effect can now be measured directly using the independently calibrated force-velocity relation [Fig.~\ref{fig1}~D and \eqref{eq:velocity}]. By monitoring how the trajectories are deflected with respect to the downward motion of the drops, we obtain $F$. 

\begin{figure}[tb]
	\centerline{\includegraphics[width=\colwidth]{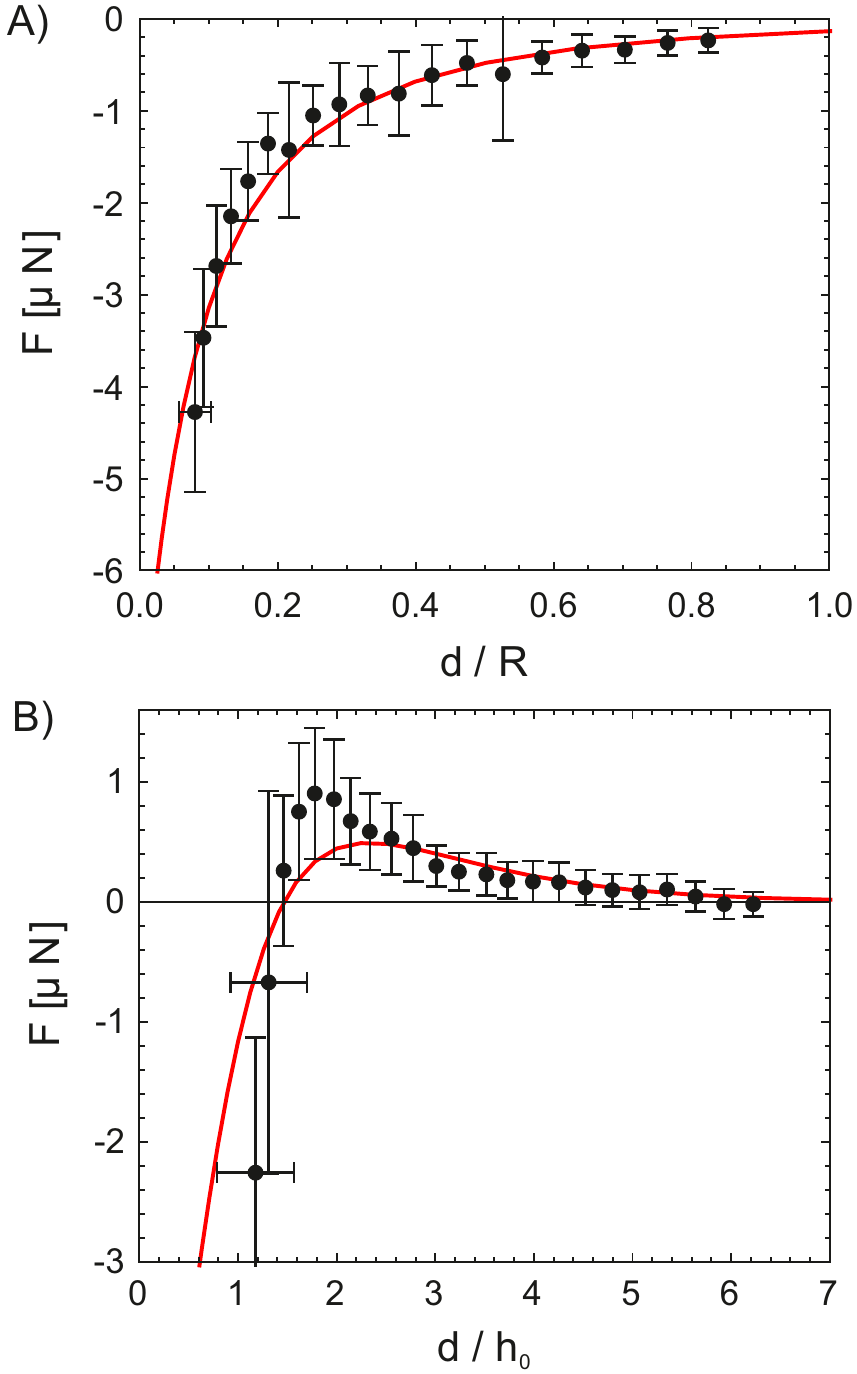}}
	\caption{Measured interaction force $F$ (symbols) as a function of their separation $d$, compared to three-dimensional theory (red lines, no adjustable parameters). A) Attraction on a thick elastic layer ($h_0 \approx 8 \mathrm{mm} \gg R \gg \ell$). B) Repulsion and attraction on a thin layer ($R \gg \ell \gtrsim h_0 \approx 40 \mathrm{\mu m}$). Each datapoint represent an average over $\sim 10$ realisations, the error bars giving the standard deviation. Measurements are based on pairs of ethylene glycol drops whose radii are in the range $R \sim 0.7 \pm 0.1 \mathrm{mm}$. The elastic substrate has a static shear modulus $0.28 \mathrm{kPa}$.}
	\label{fig2}
\end{figure}

The key result is shown in Fig.~\ref{fig2}, where we report the interaction force $F$ as a function of distance $d$. Panel A shows experimental data for the attractive force ($F<0$) between drops on thick layers (black dots), together with the theoretical prediction outlined below. Supplementary Video 1 shows an example of attractive drop-drop interaction. The attractive force is of the order of $\mu$N, which is comparable to both the capillary force-scale $\gamma R$ and the elastic force-scale $GR^2$. The force decreases for larger distance and its measurable influence was up to $d \sim R$. 

Figure~\ref{fig2}~B shows the interaction force between drops on thin layers. The dominant interaction is now repulsive ($d\gtrsim h_0$), see Supplementary Video 2. Intriguingly, we find that the interaction is not purely repulsive, but also displays an attractive range at very small distance. It is possible to access this range experimentally in case the motion of the individual drops are sufficiently closely aligned (Supplementary Video 3). The ``neutral" distance where the interaction force changes sign appears when the separation is comparable to the substrate thickness $h_0$, suggesting that the key parameter governing whether the drops attract or repel is the thickness of the gel. 

\section{Mechanism of interaction: rotation of elastic meniscus}
We explain the attraction versus repulsion of neighboring drops by computing the total free energy of drops on gel layers of different thicknesses. In contrast to the normal Cheerios effect, which involves two rigid particles, in the current experiment an additional element of complexity is present: both the droplet and the elastic substrate are deformable, and their shapes will change upon varying the distance $d$. Hence, the interaction force involves both the elastic and the surface tension contributions to the free energy that emerge from self-consistently computed shapes of the drops and elastic deformations.
\begin{figure}[tb]
	\centerline{\includegraphics[width=\colwidth]{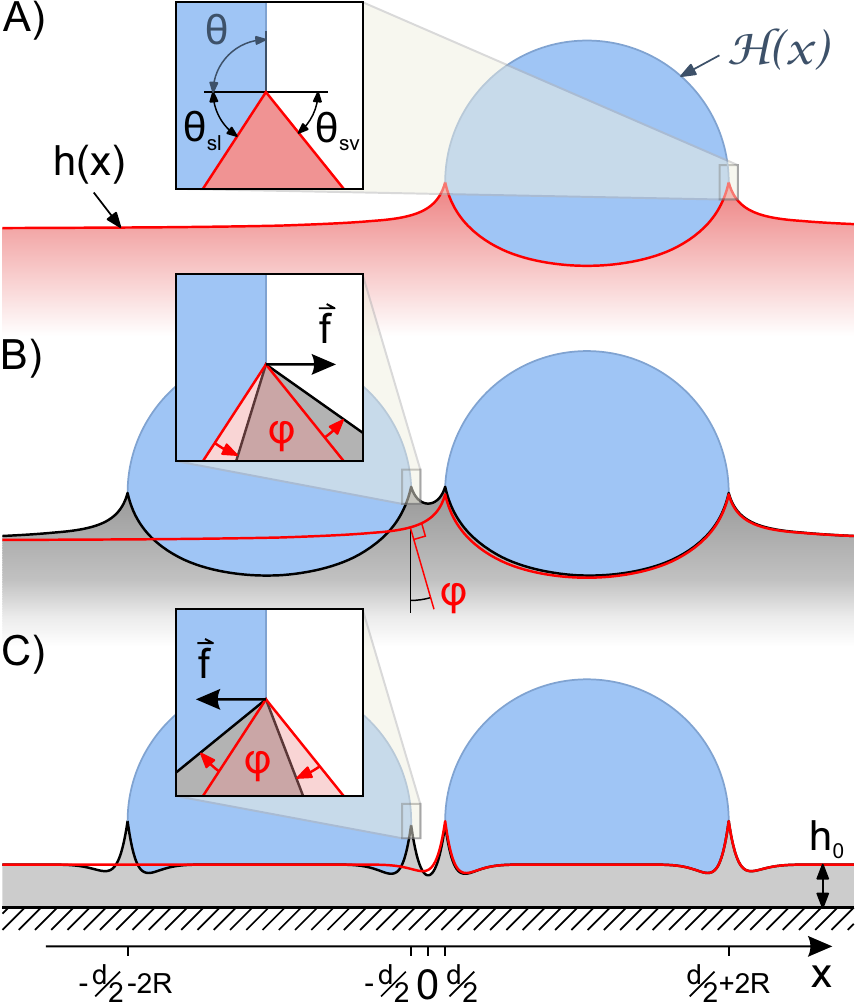}}
	\caption{Mechanism of interaction between two liquid drops on a soft solid. A) Deformation $h(x)$ induced by a single droplet on a thick substrate. The zoom near the contact line illustrates that the contact angles satisfy the Neumann condition. B) A second drop placed on a thick substrate experiences a background profile due to the deformation already induced by the drop on the right. This background profile is shown in red. As a consequence, the solid angles near the elastic meniscus rotate by an angle $\varphi$ (see zoom). This rotation perturbs the Neumann balance, yielding an attractive force $\vec{f}$. C) The single-drop profile on a thin substrate yields a non-monotonic elastic deformation. The zoom illustrates a rotation $\varphi$ in the opposite direction, leading to a repulsive interaction.}
	\label{fig:mechanism}
\end{figure}

To reveal the mechanism of interaction, we first consider two dimensional drops, for which the free energy can be written as
\begin{eqnarray}
\E[h]&=&\E_{e\ell}[h]+\int\limits_{\rm dry}\id x \, \gamma_{SV}\sqrt{1+{h'}^2}\nonumber\\
&+&\int\limits_{\rm wet}\id x  \left[ \gamma \sqrt{1+\h'^2}+ \gamma_{SL} \sqrt{1+{h'}^2}\right],
	\label{eq:tot_ener1}
\end{eqnarray}
The geometry is sketched in Fig.~\ref{fig:mechanism}~BC, and further details are given in the Supplementary Information. The elastic energy $\E_{e\ell}$ in the entire layer is a functional of the profile $h(x)$ describing the shape of the elastic solid: the functional explicitly depends on the layer thickness, and is ultimately responsible for the change from attraction to repulsion. The function $\h(x)$ represents the shape of the liquid-vapor interface. The integrals in \eqref{eq:tot_ener1} represent the interfacial energies; they depend on the surface tensions $\gamma$, $\gamma_{SL}$, $\gamma_{SV}$ associated with the liquid-vapor, solid-liquid and solid-vapor interfaces, respectively. 
Variation with respect to the contact line positions provide the relevant boundary conditions to the problem \cite{LUUKJFM14}. In order to determine the force per unit length $f$ characterizing the lateral interaction between the two-dimensional drops, the free energy \eqref{eq:tot_ener1} is minimized under the constraint that the nearest contact lines are separated by a distance $d$. This constraint is imposed by adding the term $f\;(x_i-d/2)$ to the free energy, where $x_i$ is the inner contact line position (see Supplementary Information). With this convention attractive forces correspond to $f<0$. Equivalently, $-f$ is the external force necessary to prevent the drop from moving towards each other, yielding an external work $-f \delta d$ when varying the spacing between the drops. 

The energy minimization reveals the mechanism of drop-drop interaction: the interaction force $f$ appears in the boundary condition for the contact angles,
\begin{equation}\label{eq:neumannF}
f = \gamma \cos \theta + \gamma_{SL} \cos \theta_{SL} - \gamma_{SV} \cos \theta_{SV},
\end{equation}
where the angles are defined in Fig.~\ref{fig:mechanism}. Equation \eqref{eq:neumannF} can be thought of as an ``imbalance" of the usual Neumann boundary condition. 
For a single droplet, the contact angles satisfy Neumann's law, which is \eqref{eq:neumannF} with $f=0$ (Fig.~\ref{fig:mechanism}~A). On a thick elastic layer, the overall shape of the wetting ridge is of the form \cite{Style12,LUUKJFM14}
\begin{equation}\label{eq:scaling}
h(x) \sim \frac{\gamma}{G} \, \Psi \left( \frac{x}{\gamma_s/G} \right),
\end{equation}
where the horizontal scale is set by elasto-capillary length based on the typical solid surface tension $\gamma_s$. The origin of $f$ can be understood from the principle of superposition. Due to the substrate deformation of a single drop, a second drop approaching the first one will see a surface that is locally rotated by an angle $\varphi \sim h' \sim \gamma/\gamma_s$. The elastic meniscus near the inner contact line of this approaching drop gets rotated by an angle $\varphi$  (Fig.~\ref{fig:mechanism}~B). Importantly, changes in the liquid angle $\theta$ exhibit a weaker dependence $\sim h/R \sim \gamma/(GR)$, which for large drops can be ignored. As a consequence, this meniscus rotation induces an imbalance of the surface tension forces according to \eqref{eq:neumannF}, which for small rotations yields $f \simeq \gamma \varphi$, where $\varphi$ follows from the single drop deformation \eqref{eq:scaling}. There is no resultant interaction force from the stress below the drop, which, due to deformability of the drop, provides a uniform pressure. 

The inverted Cheerios effect is therefore substantially different from the Cheerios effect between two particles floating at the surface of a liquid. Apart from the drop being deformable, we note that the energy driving the interaction is different for the two cases: while the liquid interface shape is determined by the balance between gravity and surface tension in the Cheerios effect, the solid shape is determined by elasto-capillarity in the inverted Cheerios effect. Another difference is the mechanism by which the interaction is mediated. The Cheerios effect is primarily driven by a change in gravitational potential energy which implies a vertical displacement of particles: a heavy particle slides downwards, like a bead on a string, along the deformation created by a neighboring particle \cite{Vella2005}. A similar interaction was recently discussed for rigid cylinders that deform an elastic surface due to gravity \cite{Maha2015}. In contrast, the inverted Cheerios effect discussed here does not involve gravity and can be totally ascribed to elasto-capillary tilting of the solid interfaces -- as in Fig.~\ref{fig:mechanism} -- manifesting the interaction as a force near the contact line. 

The rotation of contact angles indeed explains why the drop-drop interaction can be either attractive or repulsive. On a thick substrate, the second drop experiences solid contact angles that are rotated counter clockwise, inducing an attractive force (Fig.~\ref{fig:mechanism}~B). By contrast, on a thin substrate the elastic deformation induced by the second drop has a non-monotonic profile $h(x)$. This is due to volume conservation: the lifting of the gel near the contact line creates a depression at larger distances (Fig.~\ref{fig:mechanism}~C). At large distance, the rotation of the contact angles thus changes sign, and, accordingly, the interaction force changes from attractive to repulsive. Naturally, the relevant length scale for this phenomenon is set by the layer thickness $h_0$.

\section{Three-dimensional theory}
\begin{figure}[tb]
	\centerline{\includegraphics[width=\colwidth]{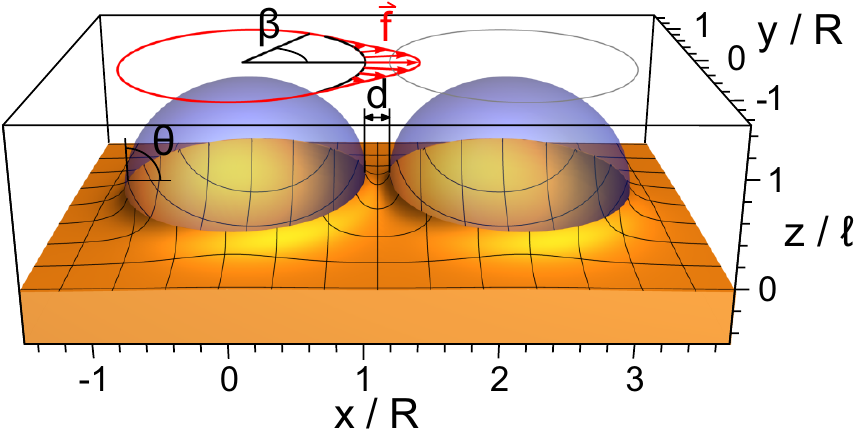}}
	\caption{Three-dimensional calculation of interface deformation for a pair of axisymmetric drops. The elasto-capillary meniscus between the two drops is clearly visible, giving a rotation of the contact angle around the drop. The total interaction force $\vec{F}$ is obtained by integration of the horizontal force $\vec{f}$ (indicated in red)  related to the imbalance of the Neumann law around the contact line. Parameter values are set to $\ell/R = 0.1$, $\gamma/\gamma_s = 1$.}
	\label{fig3}
\end{figure}

The extension of the theory to three dimensions is straightforward and allows for a quantitative comparison with the experiments. For the three dimensional case we compute the shape of the solid numerically,  by first numerically solving for the deformation field induced by a single  using an axisymmetric elastic Green function \cite{Style12}.  Adding a second drop on this deformed surface gives an intricate deformation that is shown in Fig.~\ref{fig3}. The imbalance of the Neumann law applies everywhere around the contact line: the background deformation induces a rotation of the solid contact angles around the drop. According to \eqref{eq:neumannF}, these rotations result into a distribution of force per unit length contact line $\vec{f} = f(\beta) \vec{e}_r$, where $\vec{e}_r$ is the radial direction associated with the interacting drop and $\beta$ the azimuthal angle along the contact line (Fig.~\ref{fig3}). The resultant interaction force $\vec{F}$ is obtained by integration along the contact line, as  $\vec{F}=R\int d\beta \vec f(\beta)$ (see Supplementary Information). By symmetry, this force is oriented along the line connecting the two drops. 

The interaction force obtained by the three-dimensional theory is shown as the red curves in Fig.~\ref{fig2}~AB. The theory gives an excellent description of the experimental data without adjustable parameters. The quantitative agreement indicates that the interaction mechanism is indeed caused by the rotation of the elastic meniscus.

\section{Discussion}
In summary, we have shown that liquid drops can exhibit a mutual interaction when deposited on soft surfaces. The interaction is mediated by substrate deformations, and its direction (repulsive versus attractive) can be tuned by the thickness of the layer. The measured force/distance relation is in quantitative agreement with the proposed elasto-capillary theory. The current study reveals that multiple ``pinchings'' of an elastic layer by localized tractions $\gamma$ lead to an interaction having a range comparable to $\gamma/G$. The key insight is that interaction emerges from the rotation of the elastic surface, providing a generic mechanism that should be applicable to a wide range of objects interacting on soft media. 

In biological settings, elasto-capillary interactions may play a role in cell-cell interactions, which are known to be sensitive to substrate stiffness \cite{Guo:BPJ2006}. In addition, the elastic interaction could also play a role in cell-extracellular matrix interactions, as a purely passive  force promoting aggregation between anchor points on the surface of  adhered biological cells. For example, it has been demonstrated  that a characteristic distance of about $70$~nm between topographical features enables the clustering of integrins. These transmembrane proteins are responsible for cell adhesion to the surrounding matrix, mediating the formation of strong anchor points when cells adhere to substrates \cite{huang2009,dalby2014}. Assuming that the topographical features ``pinch'' the cell with a force likely comparable to the cell's cortical tension, which takes values in the range $0.1-1\;\mathrm{mN/m}$ \cite{krieg2008,tinevez2009, fischer2014,sliogeryte2014}, and an elastic modulus of $10^3-10^4\;\mathrm{Pa}$ in the physiological range of biological tissues \cite{swift2013}, one predicts a range of interaction consistent with observations. 

More generally, substrate-mediated interactions could be dynamically programmed using the responsiveness of many gels to external stimuli (pH, temperature, electric fields). Possible applications range from fog harvesting and cooling to self-cleaning or anti-fouling surfaces, which rely on controlling drop migration and coalescence. The physical mechanisms revealed here, in combination with the fully quantitative elasto-capillary theory, paves the way for new design strategies for smart soft surfaces.

\section{Material and Methods}

The Supplementary Information provides further technical information, the variational derivation underlying Equation (3), and the numerical scheme that lead to the calculations of Fig. 2 and 4. Supplementary Videos 1, 2 and 3 show typical experiments of drop-drop interactions.

{\bf Substrate preparation~}The two prepolymer components (Dow Corning CY52-276~A and~B) were mixed in a ratio of 1.3:1 (A:B). Thick elastic layers ($\sim 8\,\mathrm{mm}$) were prepared in petri dishes (diameter $\sim 90\,\mathrm{mm}$). Thin layers ($\sim 40\,\mathrm{\mu m}$) were prepared by spin-coating the gel onto silicon wafers. Thicknesses were determined by color interferometry. See supplement for details on substrate curing \& rheology.

{\bf Determining the interaction between drops~}Droplets of Ethylene Glycol ($V \sim 0.3 - 0.8 {\mathrm \mu l}$) were pipetted onto a small region near the center of the cured substrate. The sample was then mounted
vertically so that gravity acts along the surface ($-y$ direction, compare Fig.~\ref{fig1}~A,B). The droplets were observed in transmission (thick layers) or reflection (thin layers) with collimated illumination, using a
telecenric lens (JenMetar 1x) and a digital camera (pco 1200). Images were taken every 10 s. The contours of the droplets were determined by a standard correlation technique.

At large separation, droplets move downward due to gravity. The gravitational force for each droplet is proportional to its volume. The relation between force and velocity is follows the same power law as the rheology, which was explained recently~\cite{karpitschka2015}.

Due to their different volumes/velocities, distances between droplets change. Whenever two droplets come close, their trajectories change due to their interaction. Drops on thick substrates (Figure~\ref{fig1}~C, black) attract and finally merge. On a rigid surface, these droplets would have passed by each other. Opposite holds for droplets on thin layers (red): the droplets repel each other, which prevents coalescence.

To determine the interaction forces, we first evaluate the velocity vector of each individual droplet. The droplets behave as quasi-stationary, and the total force vector acting on each droplet is aligned with its velocity vector. The magnitude of the total force is derived by the calibration shown in Fig.~\ref{fig1}~D). The interaction force is obtained by subtracting the gravitational force vector from the total force vector. Figure~\ref{fig2}~A shows data from nine individual droplet pairs, at different times and different locations on the substrate. Panel B) shows data from 18 different droplet pairs. The raw data has been averaged over distance bins, taking the standard deviation as error bar.

\begin{acknowledgments}
SK acknowledges financial support from NWO through VIDI Grant No. 11304. AP and JS acknowledge financial support from ERC (the European Research Council) Consolidator Grant No. 616918.
\end{acknowledgments}

\end{document}